\documentclass[referee]{aa} 
\usepackage{graphicx}
\usepackage{natbib}
\usepackage{txfonts}
\usepackage{ulem} 
\begin{document}
   \title{Quantification of segregation dynamics in ice mixtures}

   \author{Karin I. \"Oberg\inst{1}
          \and
		  Edith C. Fayolle\inst{1}
		  \and
		  Herma M. Cuppen\inst{1,2}
		  \and
          Ewine F. van Dishoeck\inst{2,3}
		  \and
		  Harold Linnartz\inst{1}
          }

   \institute{Raymond and Beverly Sackler Laboratory for Astrophysics, Leiden Observatory, Leiden University, P.O. Box 9513, 2300 RA Leiden, The Netherlands.\\
              \email{oberg@strw.leidenuniv.nl}
         \and
             Leiden Observatory, Leiden University, P.O. Box 9513, 2300 RA Leiden, The Netherlands
         \and
             Max-Planck-Institut f\"ur extraterrestrische Physik (MPE), Giessenbachstraat 1, 85748 Garching, Germany\\
             }

   \date{}

 
  \abstract
{The observed presence of pure CO$_2$ ice in protostellar envelopes, revealed by a double peaked 15 $\mu$m band, is often attributed to thermally induced ice segregation. The temperature required for segregation is however unknown because of lack of quantitative experimental data and this has prevented the use of ice segregation as a temperature probe. In addition, quantitative segregation studies are needed to characterize diffusion in ices, which underpins all ice dynamics and ice chemistry.}
{This study aims to quantify the segregation mechanism and barriers in different H$_2$O:CO$_2$ and H$_2$O:CO ice mixtures.}
{The investigated ice mixtures cover a range of astrophysically relevant ice thicknesses and mixture ratios. The ices are deposited at 16--50~K under (ultra-)high vacuum conditions. Segregation is then monitored, at 40--70~K in the CO$_2$ mixtures and at 23--27~K in the CO mixtures, through  infrared spectroscopy. The CO$_2$ and CO band shapes are distinctly different in pure and mixed ices and can thus be used to measure the fraction of segregated ice as a function of time. The segregation barrier is determined using rate equations and the segregation mechanism is investigated through Monte Carlo simulations.}
{Thin (8--37~ML) H$_2$O ice mixtures, containing either CO$_2$ or CO, segregate sequentially through surface processes, followed by an order of magnitude slower bulk diffusion. Thicker ices ($>$100~ML) segregate through a bulk process, which is faster than even surface segregation in thin ices. The thick ices must therefore be either more porous or segregate through a different mechanism, e.g. a phase transition, compared to the thin ices. The segregation dynamics of thin ices are reproduced qualitatively in Monte Carlo simulations of surface hopping and pair swapping. The experimentally determined surface-segregation rates follow the Ahrrenius law with a barrier of $1080\pm190$ K for H$_2$O:CO$_2$ ice mixtures and $300\pm100$ K for H$_2$O:CO mixtures. Though the barrier is constant with ice mixing ratio, the segregation rate increases with CO$_2$ concentration.}
   {Dynamical ice processes can be quantified through a combination of experiments and different model techniques and they are not scale independent as previously assumed. The derived segregation barrier for thin H$_2$O:CO$_2$ ice mixtures is used to estimate the surface segregation temperature during low-mass star formation to be $30\pm5$~K.  Both surface and bulk segregation is proposed to be a general feature of ice mixtures when the average bond strengths of the mixture constituents in pure ice exceeds the average bond strength in the ice mixture.}


   \maketitle
%

\section{Introduction}

Ices form in dark clouds through accretion of atoms and molecules onto cold (sub)micron-sized dust particles. The atoms and molecules are subsequently hydrogenated and oxygenated on the grain surface to form ices such as H$_2$O \citep{Merrill76, Tielens82, Watanabe03, Ioppolo08}. This surface formation process is efficient enough that in the densest star-forming regions up to 90\% of all molecules, except for H$_2$ exist in ice form \citep[e.g.][]{Bergin02}. Observations of dense clouds show that two of the most abundant ices, H$_2$O and CO$_2$, form already at low extinctions, while the third major ice component, CO, freezes out in the cloud core \citep{Bergin05, Pontoppidan06, Sonnentrucker08}. These results predict a bi-layered ice composition, with a bottom layer dominated by a H$_2$O:CO$_2$ mixture of $\sim$5:1 and top layer consisting of CO-rich ice. Infrared observations of the CO$_2$ ice bending mode around 15 $\mu$m towards dark clouds confirm this scenario; the CO$_2$ spectral band consists of two or more distinct components consistent with laboratory ice spectra of H$_2$O:CO$_2$ and CO:CO$_2$ ice mixtures \citep{Knez05}. In pure CO$_2$ ice spectra, the bending mode has a characteristic double peak \citep[][e.g.]{Sandford90}. This double-peak is not observed towards star-forming regions before the protostellar collapse and the turn-on of the protostar.  This agrees with current astrochemical models, where pure CO$_2$ can only form through thermal processing of previously mixed H$_2$O:CO$_2$ and CO:CO$_2$ ices. 

Heating of ice mixtures in the laboratory results in both sequential desorption, starting with the most volatile molecules, and in ice re-structuring, including ice segregation \citep[e.g.][]{Collings04, Ehrenfreund98}. In H$_2$O:CO$_2$ and CH$_3$OH:CO$_2$ ice mixtures, ice segregation is identified from the growth of pure CO$_2$ ice features, especially the characteristic 15 $\mu$m double peak. Thermal desorption of CO from CO$_2$ ice mixtures also results in the same pure CO$_2$ double-peak feature \citep{vanBroekhuizen06}. Ice heating is therefore inferred towards low and high-mass protostars wherever pure CO$_2$ ice is observed \citep{Gerakines99, Nummelin01, Pontoppidan08, Zasowski09}.  Towards most of the protostars, low- and high-mass sources alike, the majority of the CO$_2$ ice is still present in the two phases typical for dark cloud cores, suggesting a range of ice temperatures in the protostellar envelope \citep{Pontoppidan08}. The CO$_2$ spectra then contain information about the temperature structure of the protostellar surroundings. Maybe more importantly, the spectra contain information on the maximum temperature to which the icy grain mantles in the envelope have been exposed, since ice processing is irreversible. The fraction of pure CO$_2$ ice, with respect to the total CO$_2$ ice abundance towards a protostar, can therefore be used to constrain its thermal history, and its variability compared to current stellar luminosity measurements. Attempts to use this information quantitatively suffers from difficulties in determining the origin of the pure CO$_2$ ice, because of a lack of quantitative data on H$_2$O:CO$_2$ ice segregation, i.e. the segregation temperature of H$_2$O:CO$_2$ mixtures at these time scales is unknown.

Segregation is expected to occur whenever diffusion of molecules in the ice is possible and it is energetically favorable for molecules of the same kind to group together; for example H$_2$O molecules form stronger hydrogen-bonds with each other than with CO and CO$_2$. Thus, ice segregation studies provide information on diffusion barriers. These barriers are currently the most important unknowns in models of complex organic ice formation, where large molecules form through recombination of smaller molecules and radicals that diffuse in the ice \citep{Garrod08}. Ice diffusion barriers also govern selective desorption from ice mixtures, which is responsible for much of the chemical structure in protostellar envelopes. While experiments on ice chemistry and desorption provide some information on diffusion, the effects of diffusion are difficult to disentangle because of the additional dependencies of the experimental results on desorption and reaction barriers.  Segregation studies offer a comparatively `clean' environment to quantify diffusion within. 

Segregation of thick ices (0.1--10~$\mu$m or 500-50,000~monolayers) has been studied qualitatively under high-vacuum conditions by multiple groups \citep{Ehrenfreund98, Ehrenfreund99, Dartois99, Palumbo00, Bernstein05, Oberg07a}. The results differ somewhat between the different studies, but in general H$_2$O:CO$_2$ ice mixtures are observed to segregate between 60 and 75~K  at laboratory time scales for mixtures between 9:1 and 1:1. Some of the observed differences are probably due to the different mixing ratios, since \citet{Ehrenfreund99} demonstrated that segregation in CH$_3$OH:CO$_2$ ice mixtures depends on the original ice mixture composition; at 60~K only ice mixtures with ten times more CO$_2$ than CH$_3$OH segregate, at 100~K CH$_3$OH:CO$_2$ 1:3 ice mixtures segregate efficiently as well. 

Recently \citet{Hodyss08} investigated segregation in two 0.15~$\mu$m thick, H$_2$O:CO$_2$ ice mixtures (9:1 and 4:1, respectively) under high vacuum conditions. They observed an onset in segregation at 60~K, which they interpreted as resulting from the known, slow H$_2$O phase change from high density to low density amorphous ice between 38 and 68~K. The immediate segregation of an ice mixture deposited at 70~K shows, however, that segregation may also occur through diffusion.  Consistent with previous investigations they found segregation to be more efficient in the 4:1 compared to the 9:1 ice mixture. \citet{Hodyss08} also investigated the growth of segregated, pure CO$_2$ in the 4:1 ice mixture with time at three different temperatures and they observed a faster growth of the pure CO$_2$ feature, and also an increased final segregated fraction, at higher temperatures. The ice deposition temperature was important above 50~K in this study, while lack of quantified uncertainties prohibited an estimate of its significance at lower temperatures.

Building on these previous results, we aim to systematically investigate the segregation behavior of H$_2$O:CO$_2$ and H$_2$O:CO ice mixtures under astrophysically relevant conditions. The focus is on the  CO$_2$ mixtures because of their previous use as temperature tracers around protostars. The CO mixtures are mainly included to test whether the deduced segregation mechanism for H$_2$O:CO$_2$ mixtures is molecule specific or if it can be applied more generally; The H$_2$O:CO segregation temperature and mechanism are however astrophysically relevant as well since H$_2$O:CO ice is present in star-forming regions \citep{Chiar94}. For both ice mixtures, the segregation rates are measured under a range of experimental conditions, which are discussed in Section 2. The experiments are complimented by Monte Carlo simulations, introduced in Section 3, to test the theoretical outcome of different segregation mechanisms.  The results of ultra-high- and high-vacuum experiments on segregation are presented together with the deduced segregation barriers and the simulation results in Section 4. Based on the simulations and the experimentally determined segregation-rate dependencies, the possible segregation mechanisms are discussed in Section 5 followed by astrophysical implications.

\section{Experiments}

Quantitative experiments are carried out with thin ices, $<40$~monolayers (ML), under ultra-high vacuum (UHV) conditions ($\sim$10$^{-10}$ mbar) on the set-up CRYOPAD. Complimentary experiments on thick ices ($>$100~ML) are performed under high vacuum (HV) conditions ($\sim$10$^{-7}$ mbar) at a different set-up.  CRYOPAD is described in detail by \citet{Fuchs06}. The set-up is equipped with a Fourier Transform InfraRed (FTIR) spectrometer in reflection-absorption mode (called Reflection-Absorption InfraRed Spectroscopy or RAIRS), which covers  4000--800 cm$^{-1}$ with a typical spectral resolution of 1 cm$^{-1}$. The HV experiment is described in detail by \citet{Gerakines95} and is equipped with a FTIR, set up in transmission mode, which spans 4000--600 cm$^{-1}$ and is run at 1 cm$^{-1}$ resolution.

In all experiments, the ices are built up diffusively on a gold surface (UHV) or a CsI window (HV) by introducing pre-mixed gases into the chamber at the chosen deposition temperature. The gas mixtures are prepared from $^{13}$CO$_2$ and $^{13}$CO (Indugas, 98--99\% isotopic purity) and deionized H$_2$O, which is further purified by several freeze-pump-thaw cycles. Because all experiments are with $^{13}$CO and $^{13}$CO$_2$, the isotope mass is not written out explicitly in the remaining sections. The final mixture ratio is determined in each experiment using infrared spectroscopy and previously determined CO, CO$_2$ and H$_2$O transmission band strengths with $\sim$20\% uncertainty \citep{Gerakines95}.  The same band strengths are used to calculate absolute ice thicknesses, but in the thin ice experiments the band strengths are first scaled to account for the longer absorption pathway in RAIRS \citep{Oberg09b}, which results in a $\sim$50\% uncertainty in the absolute ice thickness in these experiments. Based on the results of previous spectroscopic studies on mixed ices, the H$_2$O abundance is determined from the bending mode rather than the stretching feature \citep{Oberg07a, Bouwman07}.

In each segregation experiment the ice mixture is deposited at a temperature $T_{\rm dep}$ and then quickly heated ($\sim$5~K~min$^{-1}$) to a chosen segregation experiment temperature $T_{\rm exp}$.  The relative temperatures are controlled to a fraction of a degree, while the absolute temperatures have a $\sim$2~K uncertainty. The segregation is measured by acquisition of infrared spectra during 2--4 hours; the spectral profiles of pure and mixed CO$_2$ ice are distinctly different and can thus be used to measure the amount of segregated ice. In the thin ice experiment segregation is evaluated by using the CO and CO$_2$ stretching bands and in the thick ices by combining information from the CO$_2$ stretching and bending features.

Table \ref{exps} lists the experiments carried out under ultra-high vacuum conditions. The ice-mixture thickness is varied between 8 and 37~ML to explore the thickness dependence in the ice-thickness regime accessible by RAIRS; the CO$_2$ stretching feature is linear in absorbance for the first $\sim$5--10~ML CO$_2$ ice \citep{Teolis07,Oberg09b}. At higher coverage both the spectral shape and absorbance are affected by interference. This is only true for pure CO$_2$, thus a 30~ML thick 2:1 H$_2$O:CO$_2$ ice mixture is still accessible. Most H$_2$O:CO$_2$ experiments are carried out with a  2:1 mixture, but four other mixing ratios are included as well (Exps. 1, 2, 18 and 19). The 2:1 mixture was chosen as a standard experiment to facilitate the study of segregation dependencies over a range of conditions, which is not possible for more dilute mixtures at laboratory time scales. The deposition temperature is varied between 20 and 50~K and segregation is investigated between 50 and 60~K. Two additional experiments (Exps. 16 and 17) explore the impact of fast thermal annealing and of using an ice substrate instead of depositing the ice directly on the gold surface. The ice substrate is constructed by depositing a thick layer of H$_2$O ice onto the gold surface before depositing the ice mixture. The segregating ice mixture is thus isolated from the gold surface. Five experiments on H$_2$O:CO segregation (Exps. 20--24) at different temperatures and for different ice thicknesses and mixture ratios are included to test whether the H$_2$O:CO$_2$ segregation results can be generalized to other ice compositions. Table \ref{thick_exps} lists the H$_2$O:CO$_2$ experiments carried out under high vacuum conditions, which are set up to explore temperature- and mixture-dependencies of ice segregation for thick ices between 100 and 300~ML and how this compares with the UHV thin ice experiments.

\begin{table}
\begin{center}
\caption{UHV ice-segregation experiments.}             
\label{exps}      
\centering                          
\begin{tabular}{l c c c c c c }        
\hline\hline                 
Exp. &CO$_2$/CO&H$_2$O : X &Thick. (ML) & $T_{\rm dep}$ (K) &$T_{\rm exp}$ (K) \\ 
\hline            
1 &CO$_2$ &10:1 &37 &19 &60 \\ 
2 &CO$_2$ &3:1 &36 &19 &60 \\ 
3 &CO$_2$ &2:1 &8 &19 &53 \\ 
4 &CO$_2$ &2:1 &8 &19 &58 \\ 
5 &CO$_2$ &2:1 &11 &19 &50 \\ 
6 &CO$_2$ &2:1 &11 &19 &55 \\ 
7 &CO$_2$ &2:1 &11 &19 &60 \\ 
8 &CO$_2$ &2:1 &13 &19 &50 \\ 
9 &CO$_2$ &2:1 &18 &19 &55 \\ 
10 &CO$_2$ &2:1 &18 &19 &60 \\ 
11 &CO$_2$ &2:1 &20 &19 &53 \\ 
12 &CO$_2$ &2:1 &27 &19 &56 \\ 
13 &CO$_2$ &2:1 &30 &19 &50 \\ 
14 &CO$_2$ & 2:1 &17 &40 &55 \\ 
15 &CO$_2$ &2:1 &21& 50 &55 \\ 
16$^{\rm a}$ &CO$_2$ &2:1 &10 &19 &55 \\ 
17$^{\rm b}$ &CO$_2$ &2:1 &9/75 &19 &53 \\ 
18 &CO$_2$ &1:1 &13 &19 &50 \\ 
19 &CO$_2$ &1:1 &13 &19 &55 \\ 
20 &CO &2:1 &22 &19 &27 \\ 
21 &CO &1:1 &10 &16 &27 \\ 
22 &CO &1:1 &32 &19 &23 \\ 
23 &CO &1:1 &31 &19 &25 \\ 
24 &CO &1:1 &29 &19 &27 \\ 
\hline
\end{tabular}
\end{center}
$^{\rm a}$The ice was annealed at 60~K for less than a minute before cooled back to the segregation temperature.\\ $^{\rm b}$ The H$_2$O:CO$_2$ ice mixture of 9~ML was deposited on top of 75~ML H$_2$O ice. The H$_2$O ice was deposited at 100~K to achieve a compact ice substrate and then cooled to 19~K before depositing the ice mixtures.
\end{table}

\begin{table}
\begin{center}
\caption{HV ice-segregation experiments.}             
\label{thick_exps}      
\centering                          
\begin{tabular}{l c c c c c c }        
\hline\hline                 
Exp. &H$_2$O : CO$_2$ &Thick. (ML) & $T_{\rm dep}$ (K) &$T_{\rm exp}$ (K) \\ 
\hline                 
I &4:1 &430 &15 &60\\ 
II &4:1 &510 &15 &70\\ 
III &2:1 &140 &15 &40 \\ 
IV &2:1 &160 &15 &45 \\ 
V &2:1 &240 &15 &50 \\ 
VI &2:1 &160 &15 &60 \\ 
VII &2:1 &180 &15 &70 \\ 
\hline
\end{tabular}
\end{center}
\end{table}

\section{Monte Carlo simulations}

Monte Carlo simulations are used to qualitatively investigate ice segregation through two different diffusion mechanisms. The stochastic method employed here builds on the lattice-gas Monte Carlo technique used by \citet{Los06, Los07} and \citet{Cuppen07}. In the simulations CO$_2$ and H$_2$O molecules are followed individually as a function of time, both position and environment, as they diffuse through the ice and desorb from it. The probability of a certain event in the simulation is governed by its energy barrier, which depends both on the kind of event and on the total energy difference before and after the event takes place. Exothermic events are therefore overall more likely to occur. Diffusion thus drives ice segregation when it is energetically favorable for molecules of the same kind to bind together rather than to bind with other types of molecules. The specific energy conditions for different types of diffusion is discussed further below.

In the simulation, the ice structure is modeled as a regular lattice with dimensions 25 $\times$ 25 (extended infinitely by periodic boundary conditions) $\times$ ($h + 1$), where $h$ is the ice thickness in monolayers. The lattice is initially filled randomly with the two molecular species investigated for segregation, such that the first $(h - 1)$ layers are fully filled with molecules, the $h$ layer is 75~\% filled and the last $h+1$ layer is 25~\% filled. The partial fill of the top layers simulates the expected roughness of a surface. The ratio between the two molecular species occupying the sites is a given input parameter. 

The  molecular distribution changes with time through two diffusion mechanisms, hopping of molecules into empty sites and swapping of molecules between sites, and through desorption into the gas phase. The rates for these three processes $R_{\rm h}$, $R_{\rm s}$ and $R_{\rm d}$ (where h is for hopping, s for swapping and d for desorption) are defined by 

\begin{equation}
\label{eq:hop}
R_{\rm i} = \nu_{\rm i} \times e^{-E_{\rm i}/T_{\rm exp}},
\end{equation}

\noindent where i=h, s or d,  $\nu_{\rm i}$ is the vibrational frequency of the molecule in its binding site, which is $\sim$10$^{12}$ s$^{-1}$ for physisorbed species \citep{Cuppen07}, $T_{\rm exp}$ is the temperature at which the ice mixture is kept, and $E_{\rm i}$ is a barrier height in K for process i. 

All barrier heights depend on the difference in binding energy before and after an event. In a binary mixture of molecules A and B, the total binding energy $E_{\rm b}$ depends on the number and nature of (next)-nearest neighbors ($n_{\rm nn}^{\rm A}$, $n_{\rm nn}^{\rm B}$, $n_{\rm nnn}^{\rm A}$, and $n_{\rm nnn}^{\rm B}$) surrounding the considered molecule and also on the specific molecule-to-molecule binding energies  $\varepsilon_{\rm b}^{\rm A-A}$, $\varepsilon_{\rm b}^{\rm A-B}$ and $\varepsilon_{\rm b}^{\rm B-B}$. The binding energy for next-nearest neighbors is defined to be $\frac{1}{8}$ weaker than for nearest neighbors because of the cubic geometry of the lattice. There is also an additional bulk energy term $E_{\rm bulk}$, accounting for longer range effects. The total binding energy of a molecule A is thus described by
 
 \begin{equation}
\label{eq_bin}
E_{\rm b}^{\rm A} = \varepsilon_{\rm b}^{\rm A - A}(n_{\rm nn}^{\rm A} + \frac{n_{\rm nnn}^{\rm A}}{8}) + \varepsilon_{\rm b}^{\rm B - A}(n_{\rm nn}^{\rm B} + \frac{n_{\rm nnn}^{\rm B}}{8}) + E_{\rm bulk}.
\end{equation}

\noindent The total hopping energy barrier for a molecule A into a neighboring site is defined by  
 
\begin{equation}
\label{eq_hopp}
E_h^{\rm A} = \varepsilon_{\rm h}^{\rm A} + \frac{1}{2} \Delta E^{\rm A}_{\rm b},
\end{equation}

\noindent where $\varepsilon_{\rm h}^{\rm A}$ is the barrier energy for a molecule to hop into a site where the total binding energy is the same before and after the event and $\Delta E^{\rm A}_{\rm b}$ is the difference in binding energy between the new and the old site. Hopping one site further has a barrier proportional to $2\varepsilon_{\rm h}^{\rm A}$. The total swapping energy barrier is defined by

\begin{equation}
\label{eq_swapp}
E_s^{\rm A-B} = \varepsilon_{\rm s}^{\rm A-B} + \frac{1}{2} \Delta E^{\rm A}_{\rm b} + \frac{1}{2} \Delta E_{\rm b}^{\rm B}, 
\end{equation}

\noindent where $\varepsilon_{\rm s}^{\rm A-B}$ is the barrier energy for two molecules A and B to swap into sites where the total binding energy is the same before and after the event. The desorption energy $E_{\rm d}^{\rm A}$ equals the binding energy $E_{\rm b}^{\rm A}$.

With the rates and barriers thus defined, the Monte Carlo method is used to simulate which of these processes occurs for a specific molecule through the random walk method. The simulation starts with choosing time step $\Delta t$ by picking a random number $y$ between 0 and 1 and calculating the event time $\Delta t_{\rm event}$ from

\begin{equation}
\label{eq_time}
\Delta t_{\rm event} = \frac{- {\rm ln}(y)}{R_{\rm h} + R_{\rm s} + R_{\rm d}}
\end{equation}

\noindent for all molecules in the ice. The molecule with the smallest event time is selected and its event time defines the calculation time step. This molecule will then hop, swap or desorb at $t_{\rm current} + \Delta t$ depending on the value of a second random number $z$ between 0 and $R_{\rm h} + R_{\rm s} + R_{\rm d}$. The molecular configuration in the lattice is subsequently updated to start a new calculation cycle.

In the simulation the energy barriers are chosen to represent a H$_2$O:CO$_2$ ice mixture (Table \ref{sim_val}). Since this is a qualitative model, the values are approximate. The binding energies are based on TPD experiments, except for $\varepsilon_{\rm b}^{\rm H_2O - CO_2}$, which is assigned a lower value than $\varepsilon_{\rm b}^{\rm CO_2-CO_2}$ to make CO$_2$ segregation through hopping energetically favorable within the framework of the simulation. This may be counter intuitive since CO$_2$ and H$_2$O should form a hydrogen bond, which is stronger than the van-der-Waals interactions between two CO$_2$ molecules. From TPD experiments there is however no evidence for $\varepsilon_{\rm b}^{\rm H_2O - CO_2}>\varepsilon_{\rm b}^{\rm CO_2 - CO_2}$ in ices \citep{Collings04}. This may be due to steric effects, i.e. CO$_2$ molecules may be able to bind to more molecules in a pure CO$_2$ ice compared to in a H$_2$O-rich environment, resulting in an overall stronger bond for a CO$_2$ molecule that hops from a H$_2$O dominated site to a CO$_2$ dominated site despite the lower bond strength between individual CO$_2$ molecules. This requires both more experimental and theoretical investigations. Until such data exist, segregation through hopping cannot be excluded and it is thus included in the simulations, though segregation through swapping alone is simulated as well. 

In contrast to CO$_2$ hopping, swapping only requires that the average pure ice band strengths are greater than the average band strengths in the mixture to drive segregation. In other words for the swap of a CO$_2$ molecule from an H$_2$O dominated environment with a H$_2$O molecule in a CO$_2$ dominated environment to be energetically favorable $\varepsilon_{\rm b}^{\rm H_2O - H_2O}+\varepsilon_{\rm b}^{\rm CO_2 - CO_2}>2\times\varepsilon_{\rm b}^{\rm H_2O - CO_2}$. This is expected to hold, since H$_2$O can form considerably stronger hydrogen bonds to other H$_2$O molecules compared to with CO$_2$. In addition hopping of H$_2$O drives segregation under the condition $\varepsilon_{\rm b}^{\rm H_2O - H_2O}>\varepsilon_{\rm b}^{\rm H_2O - CO_2}$ though this process was less important in driving segregation at the temperatures and barriers chosen for the simulations. Thus diffusion in the ice will drive segregation under a range of different conditions with regards to relative energy barriers, and the dominating segregation mechanism may differ between different ice mixtures.

The hopping energies are multiples of the binding energies similarly to \citet{Cuppen07} and \citet{Garrod08}, while the swapping barrier is unknown and we chose to set it to a factor three higher than the CO$_2$ hopping barrier. To investigate segregation through only hopping and only  swapping, we also run simulations with only one of the mechanisms turned on. This is done computationally by increasing the barrier height by a factor of 10 for the excluded process compared to Table \ref{sim_val}.

\begin{table}[!h]
\begin{center}
\label{sim_val}
\caption{Defined binding, hopping and swapping energies of H$_2$O and CO$_2$.}
\begin{tabular}{c c}
\hline\hline 
Barrier type$^{\rm a}$   &   Energy / K \\  
\hline
 \vspace{0.1 cm} $\varepsilon_{\rm b}^{\rm H_2O-H_2O}$     & 1000\\
\vspace{0.1 cm}  $\varepsilon_{\rm b}^{\rm CO_2-CO_2}$   & 500 \\
 \vspace{0.1 cm} $\varepsilon_{\rm b}^{\rm H_2O - CO_2}$ & 400\\
 \vspace{0.1 cm}  $\varepsilon_{\rm h}^{\rm H_2O}$     & 2400 \\
 \vspace{0.1 cm} $\varepsilon_{\rm h}^{\rm CO_2}$& 1200 \\
  \vspace{0.1 cm}    $\varepsilon_{\rm s}^{\rm H_2O-CO_2}$ & 3600 \\
 $\varepsilon_{\rm s}^{\rm CO_2-H_2O}$     & 3600 \\
\hline
\end{tabular}
\end{center}
$^{\rm a} b$ = binding, h = hopping, s = swapping.
\end{table}

\section{Results and analysis}

The results and analysis of H$_2$O:CO$_2$ segregation under UHV and HV conditions, H$_2$O:CO segregation under UHV conditions and Monte Carlo simulations are presented sequentially below. The CO$_2$ UHV experiments form the central part of this section, while the other studies present supporting information about the segregation mechanism.

\subsection{UHV CO$_2$ ice mixture experiments}

\begin{figure}
\resizebox{\hsize}{!}{\includegraphics{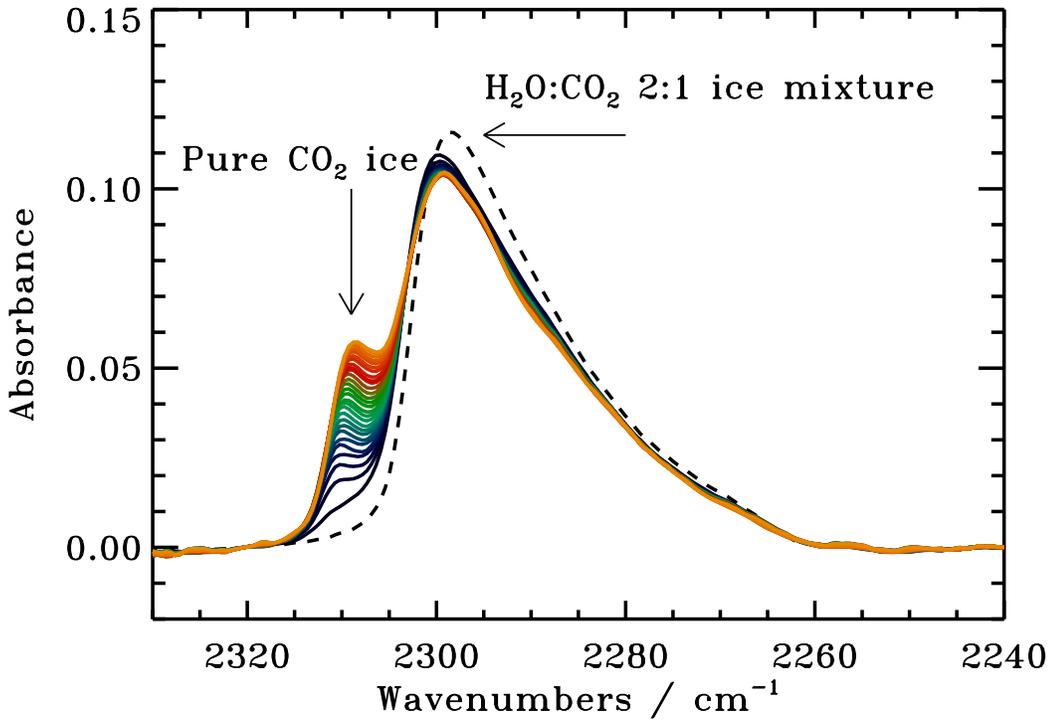}}
\caption{Spectra illustrating the changes in the CO$_2$ stretching feature as the ice segregation proceeds in a 18~ML thick 2:1 H$_2$O:CO$_2$ ice at 55 K (Exp. 9) over the course of 4 hours. The spectra are acquired every 5--15 minutes. The stretching band is shifted compared to normal CO$_2$ because $^{13}$CO$_2$ is used.}
\label{seg_sp}
\end{figure}

\begin{figure}
\resizebox{\hsize}{!}{\includegraphics{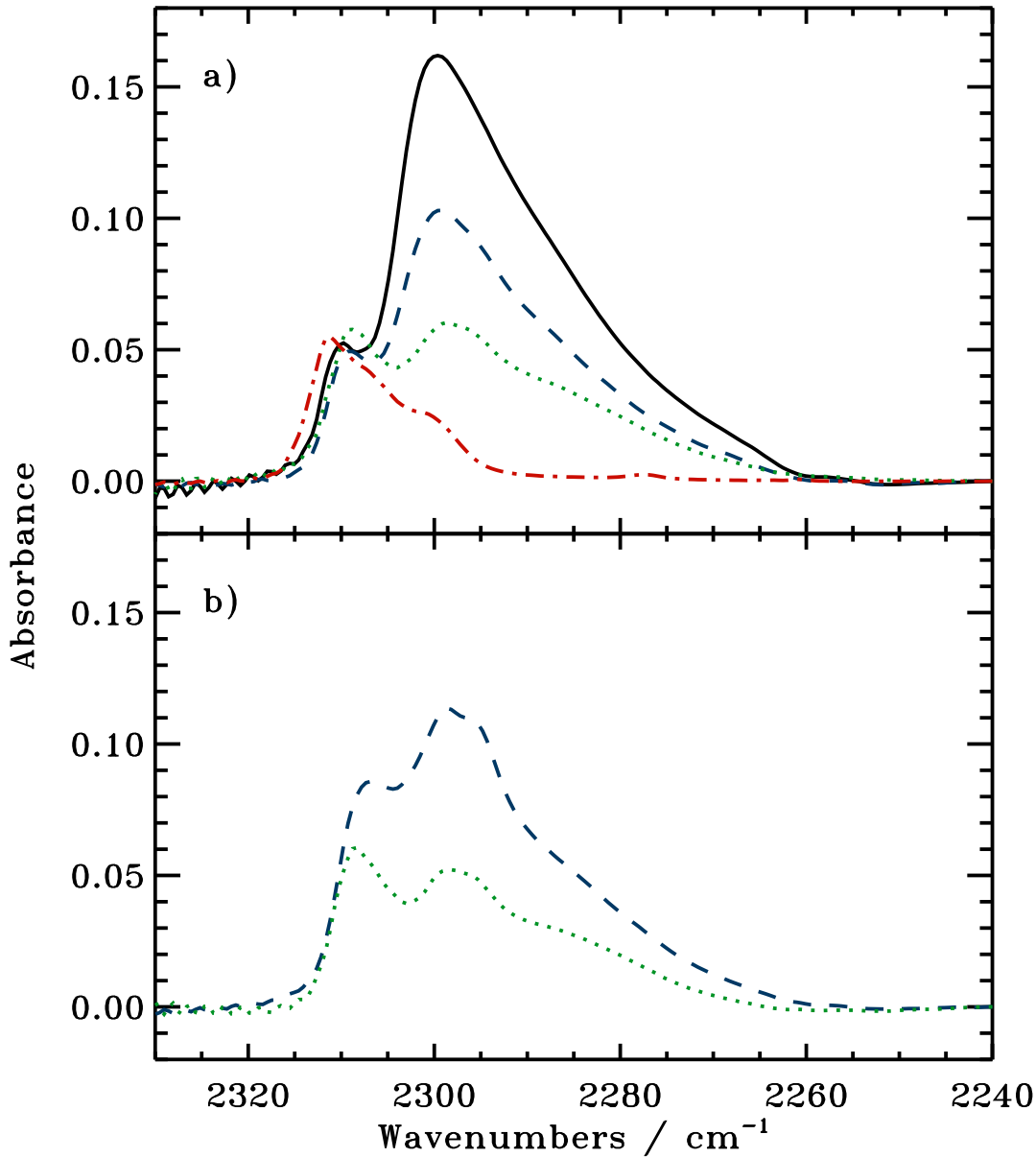}}
\caption{The spectra illustrate the level of segregation the ice reached in 2:1 H$_2$O:CO$_2$ ice mixtures of different thicknesses in a) after $\sim$3 hours at 55--56~K and in b) after $\sim$2 hours at 60~K. The ice thicknesses are 27~ML (black solid), 18~ML (blue dashed) and 11~ML (green dotted). The partly segregated ices are plotted together with a spectra of pure CO$_2$ ice at 50 K (red dashed dotted line) in panel a).}
\label{seg_level}
\end{figure}

Figure \ref{seg_sp} shows the evolution of the CO$_2$ stretching mode with time in a H$_2$O:CO$_2$ 2:1 ice mixture at 55~K during four hours. Following deposition at 20~K, the ice is first heated to 50~K, where the ice mixture spectra is acquired, and then further to 55~K. At 55~K a new feature appears within a few minutes at 2310~cm$^{-1}$, which is attributed to pure CO$_2$ ice (Fig. \ref{seg_level}). The ice segregates quickly during the first hour before the rate levels off; after four hours the segregation still proceeds slowly however. Similarly to this experiment, all investigated H$_2$O:CO$_2$ 2:1 and 1:1 ice-mixture spectra change consistently with ice segregation when the ice mixtures are heated to 50--60~K and kept at the chosen segregation temperature for a minimum of two hours. 

The early, fast segregation is inferred to be a surface process; the observed level of segregation does not depend on the initial ice thickness between 8 and 27~ML at early times for all ice temperatures, and also at late times for experiments at 55~K or colder. Figure \ref{seg_level}a demonstrates this  for 2:1 H$_2$O:CO$_2$ ice mixtures of three different thicknesses after three hours of segregation at 55~K, when the absolute amount of segregated ice is the same for all three ice mixtures within the experimental uncertainties. Segregation is somewhat thickness dependent at 60~K between 11 and 18~ML, which is indicative of a second bulk segregation process becoming efficient at this higher temperature (Fig. \ref{seg_level}b).

\begin{figure}
\resizebox{\hsize}{!}{\includegraphics{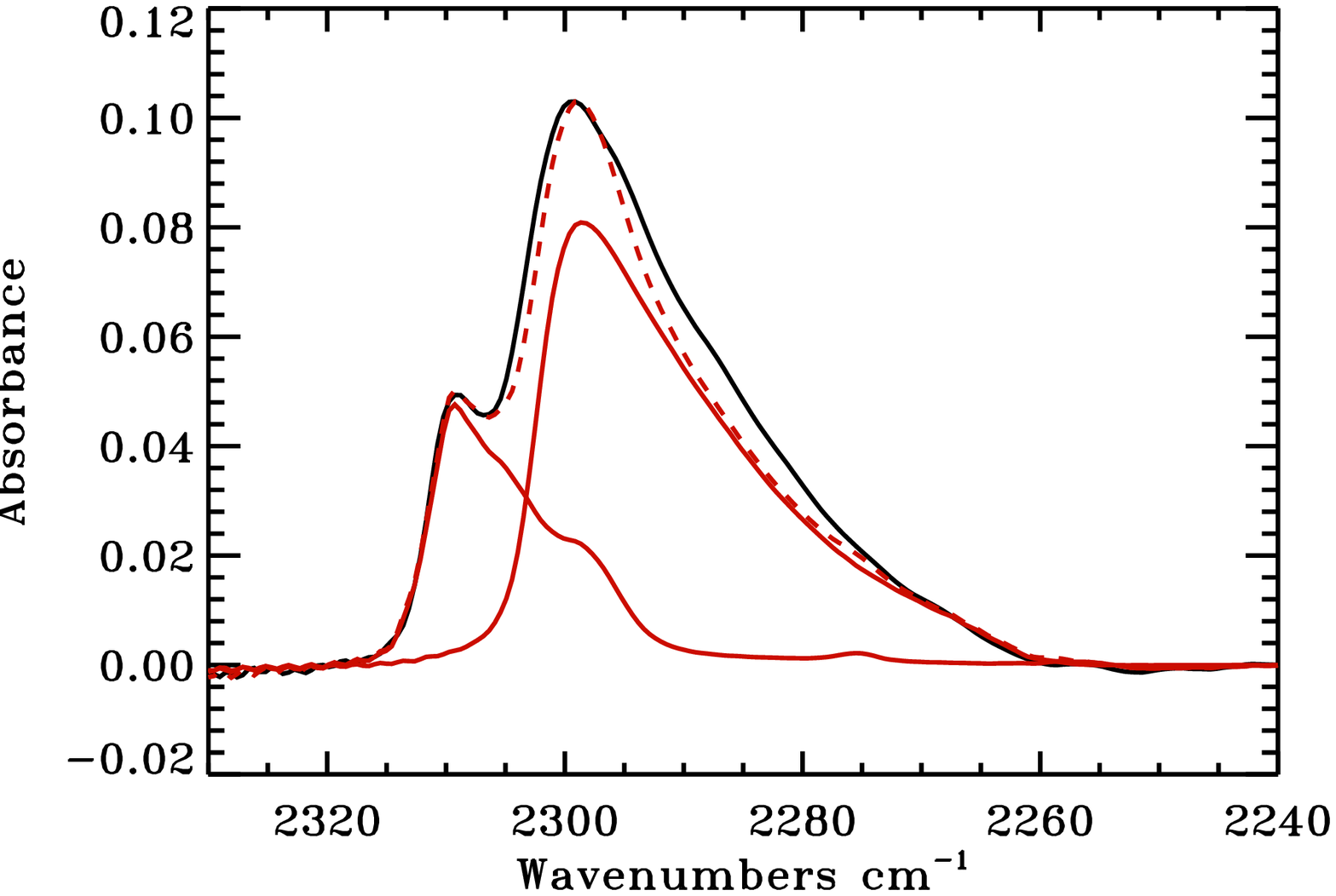}}
\caption{The spectra of an 18~ML thick H$_2$O:CO$_2$ ice with the initial composition of 2:1 (black) is fitted with a pure CO$_2$ spectra (2311 cm$^{-1}$) and a H$_2$O:CO$_2$ 2:1 mixture spectra at 50~K (red solid) after $\sim$3 hours at a segregation temperature of 55~K. The combined fit is the red dashed line.}
\label{seg_fit}
\end{figure}

\begin{figure*}
\resizebox{\hsize}{!}{\includegraphics{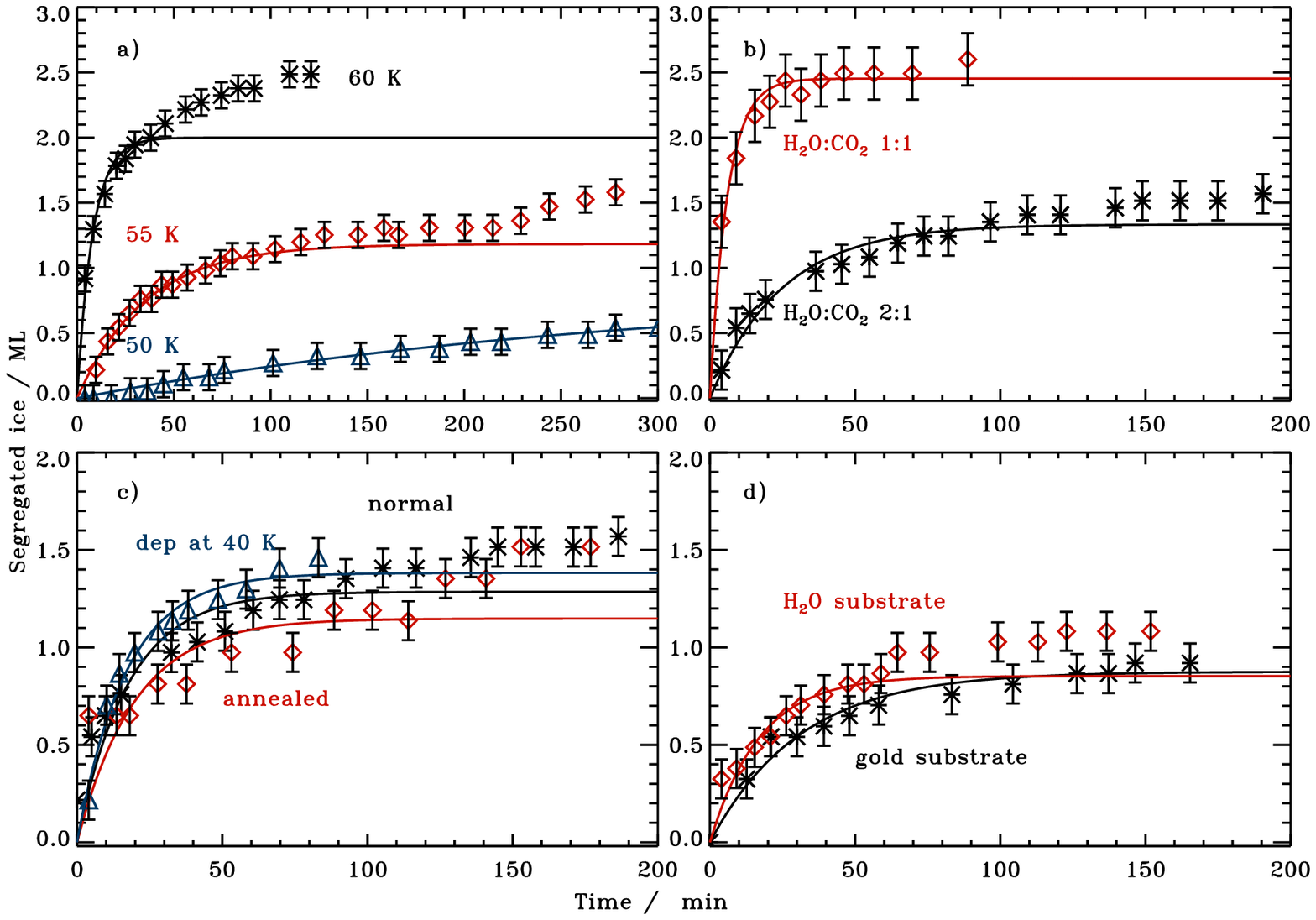}}
\caption{The amount of segregated CO$_2$ ice (in ML) as a function of time in H$_2$O:CO$_2$ ices. Panel a) shows the increasing segregation rate with temperature when comparing experiments at 50 (blue triangles), 55 (red diamonds) and 60~K (black stars) (exp. 8--10). Panel b) shows the segregation at 55~K in a 1:1 (red diamonds) and a 2:1 mixture (black stars) in two 13~ML thick ices (exp. 9 and 20). Panel c) compares the segregation at 55~K for a quickly annealed ice (red diamonds) with amorphous ice deposited at the `normal' 19~K and at 40~K (exp. 8, 14 and 16). Finally the curves in d) show the segregation at 53~K for 8-9~ML thick ices deposited directly on the gold substrate and on top of 75~ML compact H$_2$O ice respectively (exp. 3 and 17). In each experiment, the segregation during the first 40--240~min is fitted with a function $A_{\infty}(1-e^{-k_{\rm seg} t})$ (solid lines).}
\label{depends}
\end{figure*}

Quantitatively, the amount of ice that is segregated at any time is calculated by simultaneously fitting a pure CO$_2$ ice spectrum and an appropriate H$_2$O:CO$_2$ ice mixture spectrum  -- both acquired at 50~K, where segregation is still slow -- to the segregating ice using an IDL script. During the automatic fit, the template spectra are allowed to shift with at most a few wavenumbers to obtain an optimal fit as examplified in  Fig. \ref{seg_fit}. The fits are generally good except for a slight low-frequency mismatch, indicating that the structure of the mixed ice changes somewhat during heating. The resulting amount of segregated ice is plotted as a function of time in Fig. \ref{depends}a for three 13--18~ML thick 2:1 H$_2$O:CO$_2$ ice mixtures at 50, 55 and 60~K, demonstrating that the segregation rate increases with temperature. The same trend is present for both thinner and thicker ices (not shown). 

The segregation rate also depends on the ice mixing ratio as shown in Fig. \ref{depends}b for 1:1 and 2:1 H$_2$O:CO$_2$ mixtures at 55~K. The initial rate is $\sim$3 times higher in the H$_2$O:CO$_2$ 1:1 ice. This is also the case when comparing the 1:1 and 2:1 mixtures at 50~K. The upper limit on the segregation rate in the 10:1 mixture is a factor of 10 lower than the 2:1 rate. The segregation rate thus increases with a power 2--3 with the CO$_2$ concentration in the ice mixture (where the concentration is with respect to the total H$_2$O+CO$_2$ ice amount). Furthermore the steady-state segregation level is 70\% higher in the 1:1 ices compared to the 2:1 mixtures, which is comparable to the CO$_2$ concentration increase. This suggests that the same total amount of ice is available for surface segregation in all ice mixtures and that the final level of ice segregation only depends on the amount of CO$_2$ in the top available ice layers.

In contrast to temperature and composition, the segregation rate is independent of the other experimental variables, including thermal annealing, deposition temperature below 50~K and the nature of the substrate (Figs. \ref{depends}c and d). Figure \ref{depends}c shows that an ice segregating at 55~K after a 1~min warm-up to 60~K behaves as a corresponding amorphous ice at 55~K, except for the first point, which can be attributed to segregation at 60~K. The same figure shows that increasing the deposition temperature from 18 to 40~K does not affect the segregation, while at 50~K the ice segregates during deposition (not shown). Any effect by the substrate on the segregation was investigated by comparing the segregation of two thin ices (8--9~ML) at 53~K after deposition directly on the gold substrate and on top of H$_2$O ice; there is no measurable difference between the outcome of the two experiments.

The initial segregation rate is measured quantitatively in each experiment by fitting an exponential function $A_{\infty}(1-e^{-k_{\rm seg} t})$ to the first 40--240 minutes of the segregation, where $A_{\infty}$ is a steady-state amount of segregated ice, $t$ is the segregation time in seconds and $k_{\rm seg}$ is a temperature dependent rate coefficient describing the segregation process. Such a functional form is expected for a process that eventually reaches a steady-state and as demonstrated in Section 4, this is consistent with the simulations of segregation through surface diffusion.  Fig. \ref{depends} shows that the equation fits the early segregation well but at later times, the experiments and exponential fits deviate, especially for the warmest ices (Fig. \ref{depends}a, 60~K). This suggests two types of segregation mechanisms with different energy barriers, which are assigned to surface and bulk processes in Section 4. There are however too few points to fit both types of segregation in most experiments and thus we focus on the initial segregation rate alone. This is done by fitting the exponential growth function to early times, which is defined as 240~min at 50~K, 90~min at 52--56~K and 40~min above 56~K. At these times a single exponent fits the growth equally well as two exponents and the segregation is thus dominated by the fast process. To test the sensitivity of the fits to the fitting time intervals, these were varied by $\pm$50\%. This does not affect the derivation of the segregation barrier, significantly, as shown further below. The results of these fits are shown for all experiments in Table \ref{fits}. The rates are independent of ice thickness for 8--20~ML thick ices, while the thickest ice mixtures segregate somewhat slower compared to thinner ices kept at the same temperature. This decrease is barely significant, however, and it is probably due to RAIRS effects, which become important at these thicknesses  as discussed in Section 2, rather than to real differences in the segregation behavior above 20~ML. 

\begin{table}
\begin{center}
\caption{UHV ice-segregation coefficients for H$_2$O:CO$_2$ ices.}             
\label{fits}      
\centering                          
\begin{tabular}{l c c }        
\hline\hline                 
Exp. &$A_{\infty}$ / ML &$k$ / s$^{-1}$\\
\hline                 
1 &-- &--\\
 2 & 1.7$\pm$ 0.1 &$(1.4\pm0.3)\times10^{-3}$\\
 3 & 0.8$\pm$ 0.1 &$(5.4\pm2.4)\times10^{-4}$ \\
 4 & 1.2$\pm$ 0.1 &$(1.7\pm0.5)\times10^{-3}$ \\
 5 & $<$4 &$(3.1\pm6.3)\times10^{-5}$ \\
 6 & 1.3$\pm$ 0.1 &$(7.9\pm1.9)\times10^{-4}$\\
 7 & 1.7$\pm$ 0.1 &$(3.3\pm0.8)\times10^{-3}$ \\
 8 & 0.9$\pm$ 1.2 &$(5.4\pm9.6)\times10^{-5}$\\
 9 & 1.2$\pm$ 0.2 &$(4.7\pm1.5)\times10^{-4}$ \\
10 & 2.0$\pm$ 0.1 &$(2.1\pm0.4)\times10^{-3}$ \\
11 & 1.4$\pm$ 0.1 &$(4.9\pm1.0)\times10^{-4}$ \\
12 &1.3$\pm$0.4 &$(2.6\pm1.4)\times10^{-4}$ \\
13 &$<$4 &$<1.7\times10^{-4}$ \\
14 & 1.4$\pm$ 0.1 &$(9.9\pm2.1)\times10^{-4}$\\
15 & -- &-- \\
16 & 1.1$\pm$ 0.1 &$(1.1\pm0.3)\times10^{-3}$\\
17 & 1.0$\pm$ 0.1 &$(5.9\pm1.7)\times10^{-4}$\\
18 & 1.6$\pm$ 0.4 &$(4.2\pm1.8)\times10^{-4}$ \\
19 & 2.4$\pm$ 0.1 &$(2.9\pm0.4)\times10^{-3}$ \\
\hline
\end{tabular}
\end{center}
\end{table}

\begin{figure}
\resizebox{\hsize}{!}{\includegraphics{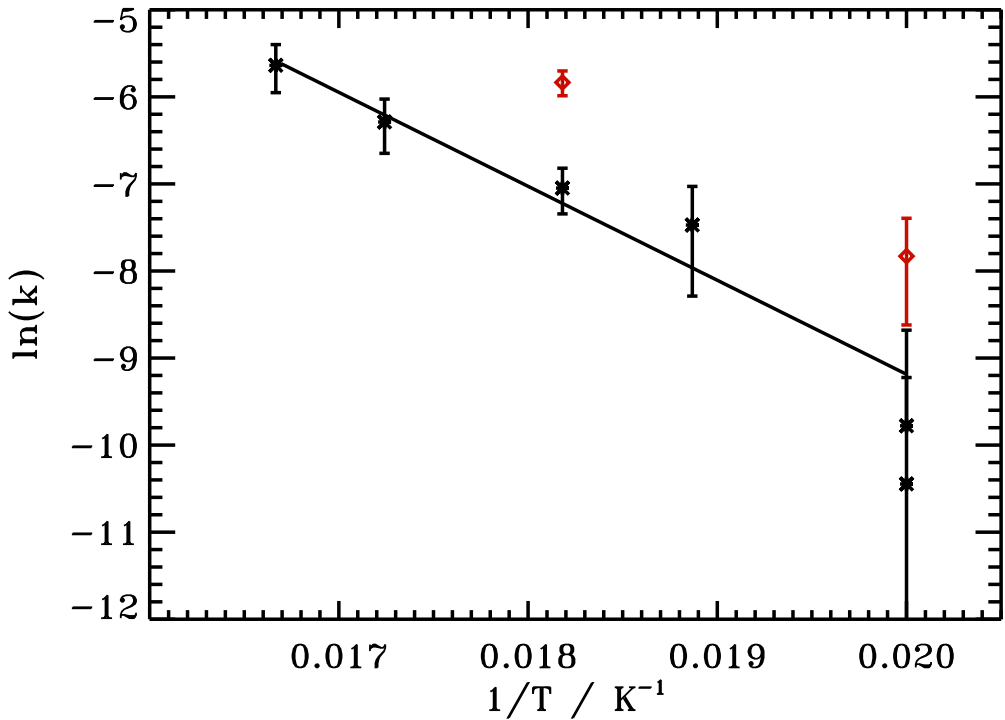}}
\caption{The natural logarithm of rate coefficients for segregation in thin ices (8--13~ML) is plotted versus the inverse of the segregation temperature for H$_2$O:CO$_2$ 2:1 (black crosses) and 1:1 mixtures (red diamonds). The black line is fitted to the 2:1 experiments.}
\label{seg_E}
\end{figure}

Assuming that the segregation-rate coefficients can be described by the Arrhenius equation, $k=\Gamma \times e^{-E_{\rm seg}/T}$, where $\Gamma$ is the pre-exponential factor, $E_{\rm seg}$ the segregation barrier in K and $T$ the ice temperature, the segregation energy can be extracted by plotting ${\rm ln}(k)$ versus $1/T$ and fitting a straight line to the points; the slope of the line is $E_{\rm seg}$. This is shown in Fig. \ref{seg_E} for the rate coefficients belonging to the 8--13~ML experiments, resulting in a segregation barrier of $1080\pm190$~K for the 2:1 and $1050\pm220$~K for the 1:1 H$_2$O:CO$_2$ ice mixtures, where the uncertainties include both experimental and fit errors. If the time intervals chosen to fit the first exponential is allowed to vary by $\pm$50\% from the optimal value, the derived barrier is instead 1200$\pm$250~K for the 2:1 mixtures. This probably overestimates the uncertainties. Still it is comforting that the derived value is well within the uncertainty of the segregation barrier derived from the optimal fitting time intervals, and the presented barrier for surface segregation is thus robust. The barriers are also not significantly different for the two mixtures, even though the rates are higher for the 1:1 ices; the pre-exponential factors amount to $2\times10^{(5\pm1)}$ for the 2:1 mixtures and $6\times10^{(5\pm1)}$ for the 1:1 mixtures.  The pre-exponential factor can be understood as a diffusion frequency  multiplied by the fraction of diffusion events that results in segregation. Its increase between the two mixing ratios is thus consistent with the increase in the number of CO$_2$ molecules available in the top layers.

Table \ref{fits} shows that the steady-state amount of segregated ice from the fast segregation mechanism is consistently between 1 and 2~ML in all experiments, except for the 1:1 mixtures as discussed above. The steady-state amount increases slightly between the 50-55~K and the 60~K experiments. This increase is however low enough that an average value over all ice temperatures and thicknesses can be used when quantifying the segregation process. Combining the information in this section, the amount of segregated ice $n_{\rm CO_2}^{\rm seg}$ in ML after a time $t$ in seconds at an ice temperature $T$ in K is well described by

\begin{eqnarray}
n_{\rm CO_2}^{\rm seg} = 1.3\pm0.7 \left(1-e^{-k_{\rm seg}t}\right)\times\left(\frac{x}{0.33}\right) {\:\rm and} \\
k_{\rm seg} =2\times10^{(5\pm1)} \times e^{-(1080\pm190)/T}\times\left(\frac{x}{0.33}\right)^{2.5\pm1},
\end{eqnarray}

\noindent where $x$ is the CO$_2$ concentration in the ice with respect to the total CO$_2$:H$_2$O ice amount (in astrophysical settings this will be dominated by the H$_2$O ice fraction). At 60~K the second, bulk segregation rate is an order of magnitude lower than the initial surface segregation rate. There are too few measurements to quantify the bulk segregation further, but this difference suggests that the barrier to segregate in the bulk is more than a factor of two larger than for the surface segregation evaluated here. In Section 4.4 this bulk segregation is simulated through molecular pair swapping with a three times higher energy barrier compared to CO$_2$ surface hopping.

\subsection{HV CO$_2$ ice mixture experiments}

\begin{figure}
\resizebox{\hsize}{!}{\includegraphics{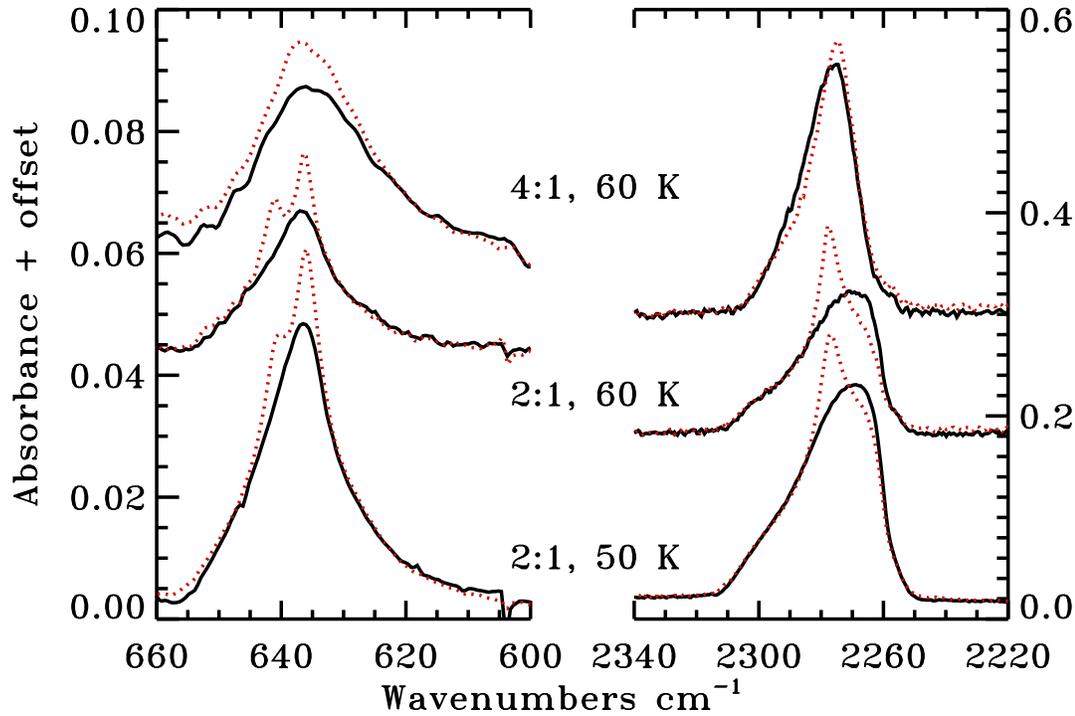}}
\caption{CO$_2$ bending ($\sim$635 cm$^{-1}$) and stretching ($\sim$2270 cm$^{-1}$) spectral features at the onset (black solid) and after 2 hours of heating at the segregation temperature (red dashed) for three 160--510 ML thick H$_2$O:CO$_2$ ice mixture experiments.}
\label{spec_thick}
\end{figure}

\begin{figure}
\resizebox{\hsize}{!}{\includegraphics{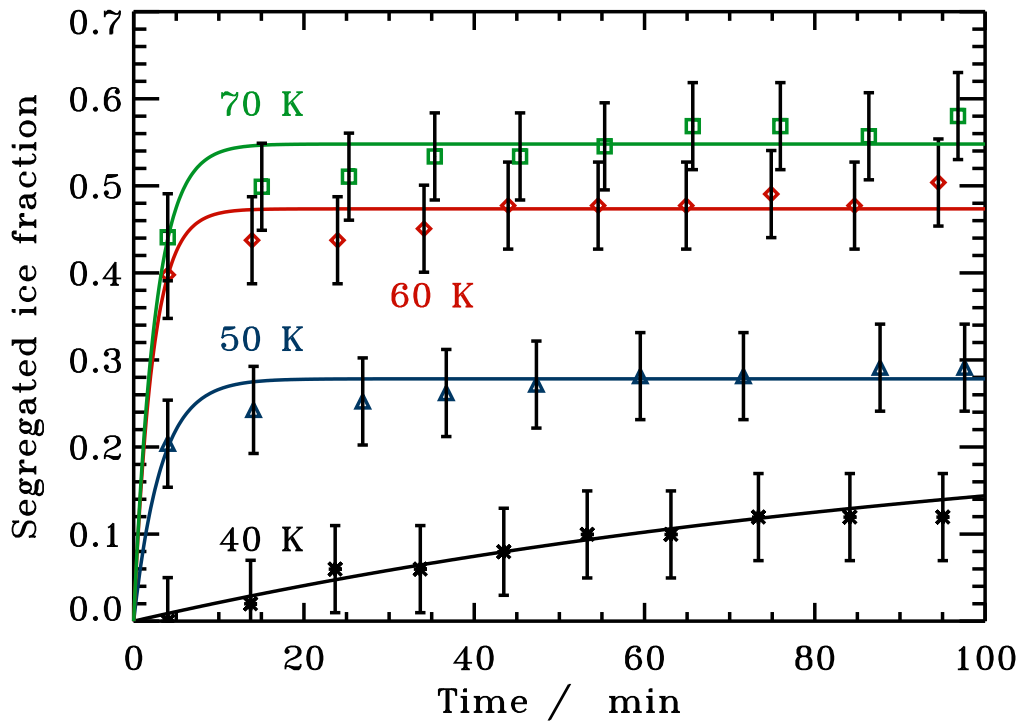}}
\caption{The segregated ice fraction, with respect to the total CO$_2$ ice amount, as a function of time at 40 (black stars), 50 (blue triangles), 60 (red diamonds) and  70~K (green circles) for 140--240~ML thick H$_2$O:CO$_2$ 2:1 ices. The solid lines show the exponential fits from which the rate coefficient is derived for each experiment.}
\label{seg_thick}
\end{figure}

Transmission spectroscopy in the  HV set-up allows for the investigation of segregation in an order of magnitude thicker ice than is possible in CRYOPAD. Figure \ref{spec_thick} shows the segregation of H$_2$O:CO$_2$ 2:1 mixtures through spectroscopy of both the CO$_2$ bending and stretching modes. Some ice rearrangement is also visible in the 4:1 mixture but there is no clear indication of segregation into two different ice phases. Thus, similarly to the thin ice experiments, the segregation rate depends on the ice mixture ratios. 

The amount of segregated ice at each time step is derived using the same procedure for the thick ices as described above for the thin ices. Figure \ref{seg_thick} shows the fraction (not the number of ML) of the ice mixture  that is segregated as a function of time at segregation temperatures between 40 and 70~K. Above 50~K the segregation of the $>$140~ML ices reaches 50\%, from which we infer, in agreement with previous studies, that segregation is a bulk process for thick ices deposited under high-vacuum conditions. It also shows however that a large fraction of the ice is protected from segregation at all ice temperatures, though this fraction is smaller than in the thin ice experiments that are dominated by surface diffusion. The segregated amount of ice as a function of time is fit to an exponential function in the low-temperature experiments -- above 45~K, segregation is too fast for a reliable rate determination, i.e. the final segregated amount is reached within a couple of measurements. In contrast to thin ices the segregation can be fitted with a single exponential function for all temperatures and the determined steady-state amount of segregated ice depends on the ice temperature; the segregated fraction varies between 20\% and 55\% at the investigated temperatures.  The resulting bulk segregation rates at 40 and 45~K are $(2\pm6)\times10^{-4}$ and $(5\pm3)\times10^{-3}$ s$^{-1}$, respectively. Since the rates describe the segregation of a fraction of the ice rather than a number of monolayers they cannot be directly compared to the thin ice surface segregation rates -- it is, however, clear that steady state is reached in a shorter time in the thick ice experiments than in the thin ices. The derived barrier from the 40 and 45~K experiments is $1270\pm560$~K, which is consistent with the thin ice segregation barrier of 1080$\pm$190~K. The thick ice segregation, while more efficient, thus does not necessarily proceed through a separate mechanism.

\subsection{UHV CO ice mixture experiments}

\begin{figure}
\resizebox{\hsize}{!}{\includegraphics{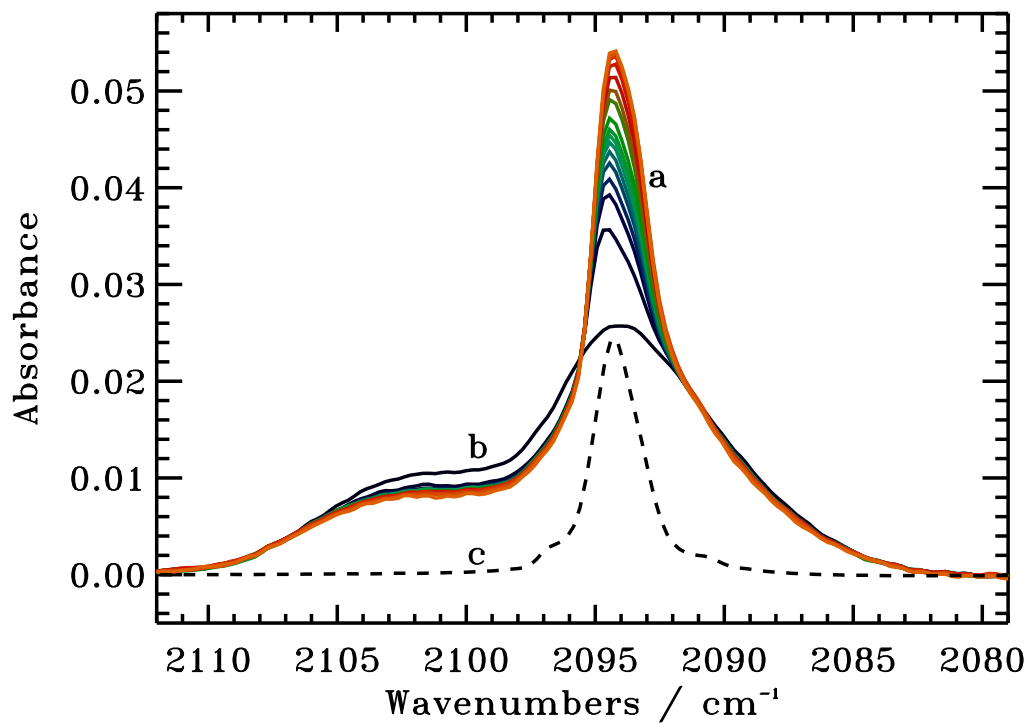}}
\caption{Ice segregation in a 10~ML H$_2$O:CO mixture observed through a) the changes in the CO stretching feature with time (from blue to orange), plotted together with b) the original ice mixture spectra (black solid) and c) a pure CO ice spectra (black dashed). The spectra are off-set compared to the normal isotope band positions, because $^{13}$CO is used.}
\label{COspec}
\end{figure}

Because of the high volatility of CO, H$_2$O:CO  segregation is investigated between 23 and 27~K. This is below the temperature range for the H$_2$O ice transition between high and low density amorphous phases, which occurs gradually between 38 and 68~K \citep{Jenniskens94}. The temperatures investigated for H$_2$O:CO and H$_2$O:CO$_2$ segregation are similar in terms of \% of the CO and CO$_2$ desorption temperature; CO desorbs at $\sim$30~K and CO$_2$ at $\sim$70~K from a H$_2$O ice substrate. Hence, if segregation occurs through a simple diffusion process with barriers proportional to the binding energy in CO and CO$_2$ ice mixtures, this is the predicted range of temperatures where the H$_2$O:CO ice should segregate. Figure \ref{COspec} shows the H$_2$O:CO ice evolution when quickly heated from 16 to 25~K and then maintained at 25~K for four hours. The new band appearing around 2094 cm$^{-1}$ at 25~K has the same width and a comparable position to a pure CO ice feature and this is interpreted as resulting from ice segregation. Segregation thus occurs and it occurs only in a part of the ice, similarly to the CO$_2$ mixtures. 

\begin{figure}
\resizebox{\hsize}{!}{\includegraphics{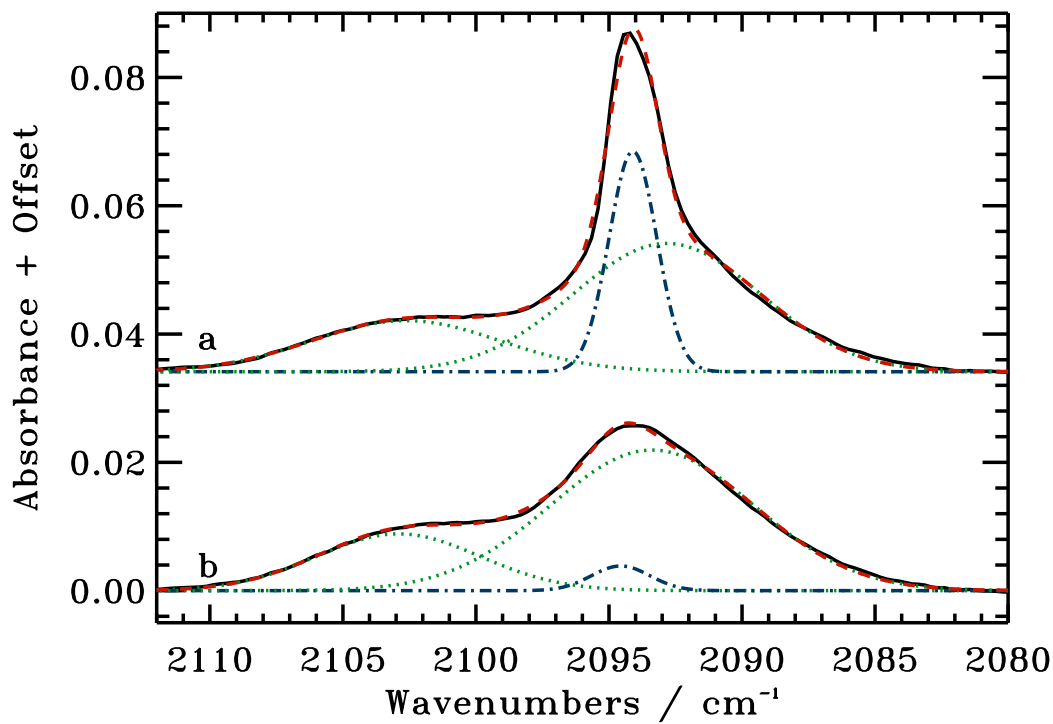}}
\caption{The resulting fit of three Gaussians to a) a partially segregated 10~ML H$_2$O:CO ice at 25~K and b) the original ice mixture at 16~K. The green dotted Gaussians correspond to mixed ice features, the blue dashed-dotted to pure CO ice and the combined fit is plotted in red dashed lines.}
\label{COfit}
\end{figure}

The segregation is quantified by fitting three Gaussian's to the spectral feature, two belonging to CO ice in a H$_2$O mixture and one corresponding to the pure CO ice band (Fig. \ref{COfit}). The two mixed-ice Gaussians have been observed previously in several studies and are attributed to two different CO-H$_2$O interactions \citep{Collings03}. The Gaussian parameters are free variables during the fit except for the FWHM of the pure CO Gaussian, which is constrained to be within 0.5 cm$^{-1}$ of the pure CO ice value in \citet{Bouwman07}. The resulting Gaussian parameters for all H$_2$O:CO experiments are reported in Table \ref{CO_gauss}. The widths of Gaussians 1 and 3 are similar to those used to fit the same H$_2$O:CO ice-mixture feature by \citet{Bouwman07}. Figure \ref{COfit} shows that the spectral fit cannot exclude a few \%  of ice segregation at deposition. This does not affect the results since the reported level of segregation during the experiment is calculated from the increase of Gaussian 2 compared to deposition. 

\begin{table}[!h]
\centering
\caption{~Parameters for the three Gaussians used to fit the CO feature.}
\centering
\begin{tabular}{c c c c}
\hline  \hline
Gaussian   &   Peak position (cm$^{-1}$)      & FHWM (cm$^{-1}$)      \\  
\hline
1         &2103.0--2102.4         & 6.1--6.4     \\
2         &2094.6--2093.8          &2.1-2.5   \\
3          &2093.4--2092.7        & 8.4--8.7   \\
\hline
\end{tabular}
\label{CO_gauss}
\end{table}

\begin{figure}
\resizebox{\hsize}{!}{\includegraphics{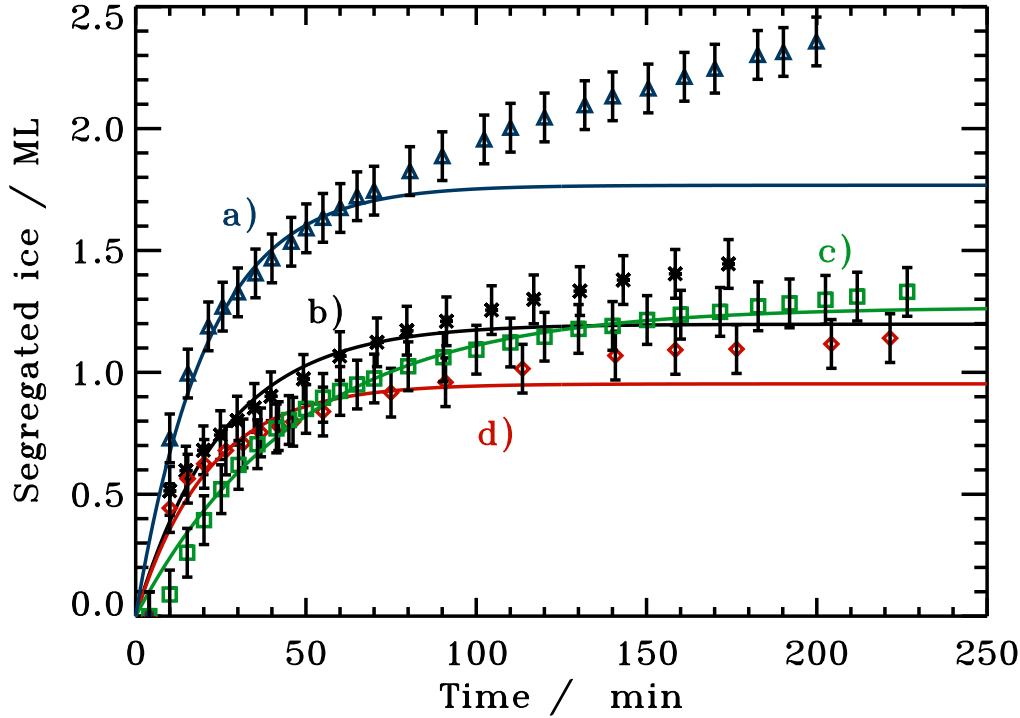}}
\caption{The amount of segregated ice as a function of time in four H$_2$O:CO mixtures: a) 1:1, 27~K, 29~ML, b) 2:1, 27~K, 22~ML, c)  1:1, 25~K, 31~ML and d) 1:1, 27~K, 10~ML.}
\label{COrates}
\end{figure}

Using the Gaussian fits, Fig. \ref{COrates} shows the time evolution of the amount of segregated ice in the different CO ice mixtures, except for the 23~K ice where no segregation is observed. The segregation time series have the same shape as the CO$_2$ segregation curves, with a fast initial segregation resulting in $<$2~ML segregated ice, followed by a slower segregation process at the highest temperature. The ice composition dependence seems weaker, but is still present, compared to the CO$_2$ case. The segregation at 27~K is thickness dependent, which suggest that bulk diffusion has already become important at this temperature, though the initial rate should still be dominated by surface processes. Comparing the early-time rates of the 1:1 mixtures at 25 and 27~K results in an approximate barrier for the initial H$_2$O:CO segregation of $300\pm100$~K, using the same fitting technique as for the CO$_2$ mixtures. This is significantly lower than the H$_2$O:CO$_2$ segregation barrier, as would be expected for the more volatile CO ice. 

\subsection{Monte Carlo simulations}

\begin{figure}
\resizebox{\hsize}{!}{\includegraphics{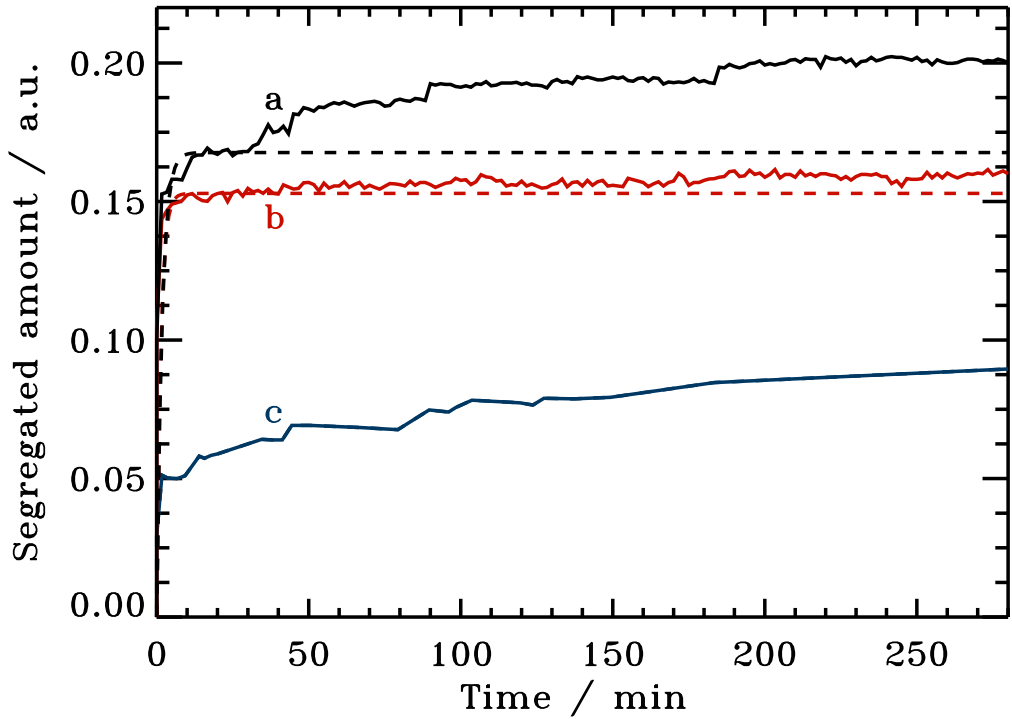}}
\caption{Simulations of ice segregation in a H$_2$O:CO$_2$ 1:1 mixture at 55~K due to a) swapping and hopping, b) only hopping, and c) only swapping. The dashed lines show exponential fits to the first 30~min.}
\label{sim_mech}
\end{figure}

This section presents the results from six different ice segregation simulations, investigating the importance of swapping and hopping, and the effect of temperature on the segregation behavior. Figure \ref{sim_mech} shows the simulated H$_2$O:CO$_2$ ice segregation in a 5~ML thick 1:1 mixture using the Monte Carlo technique and inputs described in Section 3. The segregated ice fraction is quantified through an increase in the number of nearest neighbors of the same species, which is not directly comparable to the spectroscopic segregation observed in the experiments, though the two are linked. The simulated amount of segregated ice is thus presented in arbitrary units. With the specified energy barriers, segregation at 55~K occurs through a combination of hopping and swapping; removing one of the mechanisms decreases the total segregated fraction (Fig. \ref{sim_mech}). 

The segregation through hopping is well defined by an exponential function, i.e. it is a single process approaching steady-state, while segregation through swapping, and segregation through swapping + hopping are both not. Instead segregation through swapping is initially fast and then continues to proceed slowly until the end of the simulation, long after the segregation through hopping has reached steady-state. The initial fast swapping is due to swapping with surface molecules, which explains that segregation through hopping and swapping is not additive. It also explains why segregation through only swapping cannot be fitted by a single exponential function, since it proceeds through two different processes, surface and bulk swapping. Of the three curves the swapping+hopping one is qualitatively most similar to the 55 and 60~K CO$_2$ UHV experiments and hopping alone cannot explain the non-exponential behavior of the experimental segregation curves during late times. It is however important that swapping alone may be able to reproduce the segregation behavior observed in the experiments and this implies that the CO$_2$-CO$_2$ binding energy is not required to be higher than the CO$_2$-H$_2$O one for segregation to occur through a combination of surface and bulk diffusion.

\begin{figure}
\resizebox{\hsize}{!}{\includegraphics{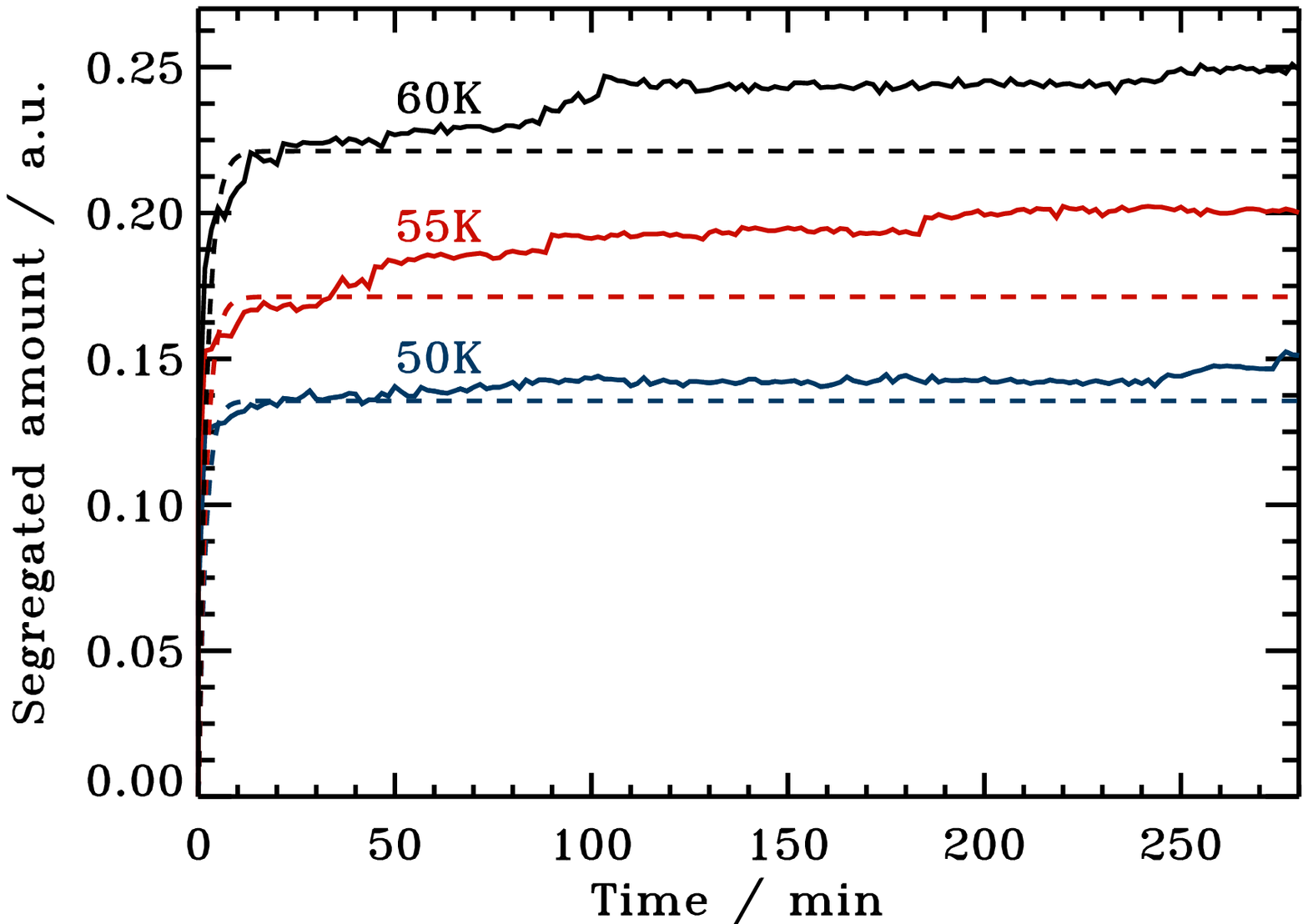}}
\caption{Simulations of ice segregation in a H$_2$O:CO$_2$ 1:1 mixture at 50, 55 and 60~K due to a combination of swapping and hopping. The dashed lines show exponential fits to the first 30~min.}
\label{sim_temp}
\end{figure}

The segregation rate increases with segregation temperature in the simulation when both hopping and swapping are included, as is observed in the experiments. Figure \ref{sim_temp} also shows that the curves at higher temperatures, i.e. 55 and 60~K, diverges more from an exponential growth compared to the 50~K simulation, suggesting an increasing importance of bulk-swapping with temperature. This is also consistent with the experimental results (Fig. \ref{depends}a). The segregated fraction in the simulations increases more clearly with temperature than in the experiments, suggesting that the bulk process, swapping of bulk molecules, may be somewhat too efficient in the simulations.

\section{Discussion}

Following a brief comparison with previous experiments, this section combines the results from  (U)HV H$_2$O:CO$_2$ and H$_2$O:CO segregation experiments and  Monte Carlo simulations to constrain the segregation mechanics and the meaning of the empirically determined segregation rates and barriers. This information is then used to discuss the astrophysical implications of quantifying the segregation of binary ice mixtures.

\subsection{Comparison with previous experiments}

Previous studies have focused on segregation in thick ices ($>$100~ML) under HV conditions. Of these studies, \citet{Hodyss08} carried out the most similar experiments to our HV ones and thus we only compare these two sets of experiments. \citet{Hodyss08} reported an increase in the segregated ice fraction and in the segregation rate between 55 and 70~K for a H$_2$O:CO$_2$  4:1 mixture, which is comparable to what we observe between 40 and 60~K for a 2:1 mixture.  The lower temperatures required in our experiment to achieve the same result are consistent with the observed dependence on composition for segregation in both studies. Similarly to \citet{Hodyss08}, we do not observe any dependence on the ice deposition temperature in our thin ice experiments, as long as the deposition temperature is lower than the lowest temperature for which segregation is observed. Qualitatively, the studies also agree on the segregation dependence on composition. A quantitative comparison is not possible since such data have not been reported previously.

\subsection{Segregation mechanisms}

\begin{figure}
\resizebox{\hsize}{!}{\includegraphics{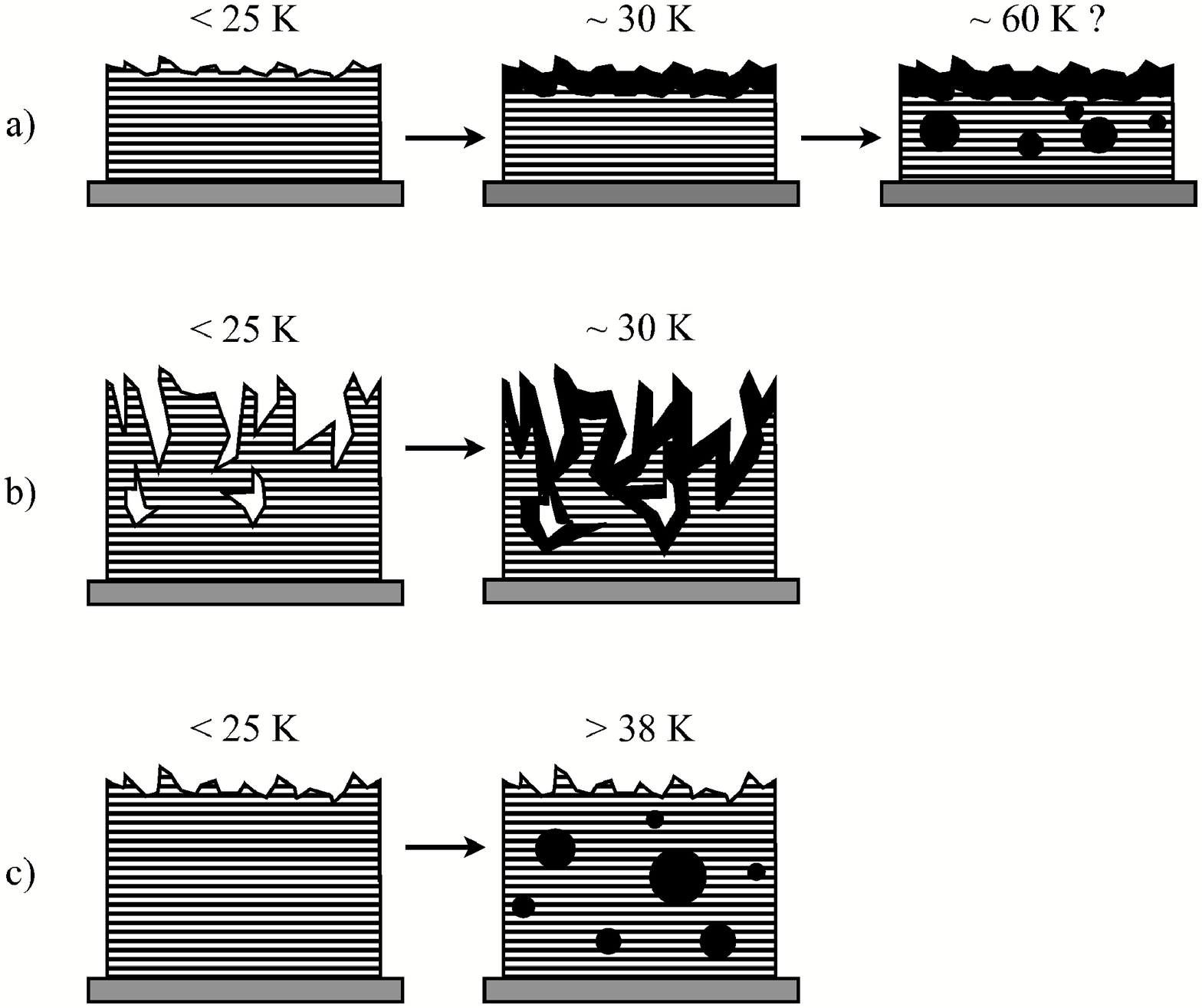}}
\caption{The segregation mechanism in a) ices below 40~ML together with the two proposed segregation mechanisms for thick ices ($>$100~ML), through b) internal surface segregation and c) phase transitions. Black indicates segregated CO$_2$ ice and stripes the H$_2$O:CO$_2$ ice mixture. Each step is marked with the approximate temperature at which the mechanism becomes important at the time-scales of low-mass star formation, as discussed in Section 5.3.}
\label{fig:cartoon}
\end{figure}

The possible segregation mechanisms for thin and thick ice mixtures are summarized in Fig. \ref{fig:cartoon}. Thin binary H$_2$O:CO$_2$ ice mixtures, 8--37~ML, segregate differently compared to thick, $>$100~ML, ices. This implies either a thickness dependent segregation mechanism  or a different ice structure above 100~ML compared to $<$40~ML. Segregation in thin ices is qualitatively reproduced by Monte Carlo simulations including surface hopping, surface swapping and the slower molecular swapping throughout the bulk of the ice. If thicker ices are more porous because of cracking, hopping and surface swapping may be possible throughout the ice resulting in the observed fast bulk segregation. Previous studies, including \citet{Hodyss08}, instead suggest that segregation is the product of a H$_2$O phase transition around the segregation temperature. Phase transitions may require a certain ice thickness to become effective and thus may only be important for thicker ices; the H$_2$O:CO results show that phase transitions are not responsible for segregation in thin ices. The similar segregation barrier in thick ices and thin ices suggests that the mechanism is the same at all thicknesses and that the rate differences have a structural origin. The shorter times required to reach steady-state in thick compared to thin ices indicates, however, the opposite. More thick ice experiments containing molecules with different segregation temperatures are thus required to differentiate between these two segregation scenarios.  In either case, these different segregation processes in thin and thick ices show that  ice dynamics, including ice chemistry, is not necessarily thickness independent as has been previously assumed in the astrochemical literature.

The segregation growth curves of thin H$_2$O:CO$_2$ and H$_2$O:CO ices have the same shape with an initial exponential growth up to 1--2~ML, followed by a slower growth at longer time scales. The segregation temperatures and inferred barriers differ by a factor 2--3. The desorption/binding energies of pure CO$_2$ and CO ices are similarly a factor of 2--3 different \citep{Collings04}. The derived segregation barriers for the two types of mixtures are thus consistent with segregation through molecular hopping and swapping, with hopping and swapping barriers proportional to the desorption/binding energy of the most volatile mixture component. The similar ratio between the surface segregation barrier and desorption energies in the two types of ice mixtures, suggests that ice surface dynamics can be approximately parameterized based on desorption energies of pure ices alone. 

Physically, it is not obvious from the simulations whether this initial surface segregation is dominated by surface hopping or surface swapping or whether it is even the same in both the H$_2$O:CO$_2$ and H$_2$O:CO mixtures. Swapping is certainly required to explain the long-time growth of the segregated ice fraction. Depending on the relative barriers for hopping and swapping and the different binding energies in the ice, it may also be responsible for at least a part of the surface segregation as shown in Fig. \ref{sim_mech}. The similar behavior of the H$_2$O:CO$_2$ H$_2$O:CO ices actually suggests that surface swapping rather than hopping drives surface segregation in both ices, since there is experimental evidence for a higher H$_2$O-CO binding energy compared to the CO-CO one in ices \citep{Collings03} and surface hopping can only drive segregation if the hopping molecules forms stronger bonds with its own kind. Quantitative H$_2$O:CO$_2$ desorption data is however required to exclude surface segregation through CO$_2$ hopping. A more quantitative simulation study of ice segregation is currently being pursued, which will address this question (Cuppen et al. in prep.). Until then, Eqs. 6 and 7 describe a non-specified surface segregation, while bulk segregation is approximately an order of magnitude slower in thin ices. From the experimental studies and the simulations, segregation is expected to occur in all ices, through swapping, where the binding energies of A-A + B-B are greater than $2\times$A-B. The segregation barrier is expected to be mixture dependent, but related to the pure ice binding energies.

\subsection{Astrophysical implications}

Nature is `non-cooperative' in that the typical astrophysical ice thicknesses of $<$100~ML are not guaranteed to be in the thin ice regime where hopping and swapping alone are responsible for ice segregation. Yet, since 100~ML is the upper limit assuming spherical grains, most astrophysical H$_2$O:CO$_2$ ice mixtures are probably closer to the thin ice regime than the thick ice one. Therefore this section will discuss ice segregation during star formation under the assumption that this can be described by the same equations as thin ice surface segregation.

During star formation ices are heated in the collapsing envelope to a certain temperature, depending on their distance from the protostar. The amount of time they spend at these elevated temperatures before desorbing depends on the infall rate that transports ices from the outer to the inner parts of the protostellar envelope. This time scale $\tau_{\rm crit}$ for ice heating in the envelope during low-mass star formation thus depends on several different parameters of which the stellar luminosity is the most important one and will vary between different objects. Building on Eq. 7, the characteristic temperature $T_{\rm seg}$ at which segregation is important can be calculated as a function of this heating time scale $\tau_{\rm crit}=k^{-1}$ for a known ice composition

\begin{equation}
T_{\rm seg} = \frac{1080\pm190}{{\rm log}\left(2\times10^{5\pm1}\times\left(\frac{X}{0.33}\right)^{2.5\pm1}\times\tau_{\rm crit} \right)}.
\end{equation}

\noindent where $x$ is the CO$_2$ concentration in the H$_2$O:CO$_2$ ice phase, which is set to 0.16 from the 5:1 ice mixture observed by \citet{Pontoppidan08}. \citet{Pontoppidan08} also calculated that under the assumption of free-fall accretion, a dust grain is heated to 25--50~K by a solar-mass protostar for $\sim$4000 years, while in a recent infall model it is at least a factor of 5 longer \citep{Visser09}. Taking the lower limit 4000 yrs time scale for segregation, $T_{\rm seg}$ is 30$\pm5$~K. This is low compared to previous estimates, of 50--100~K, and comparable to the temperature for CO distillation from CO:CO$_2$ ice mixtures, which is a competitive mechanism to form pure CO$_2$ ice. With this segregation temperature, segregation of H$_2$O:CO$_2$ ice is no longer excluded as a source for pure CO$_2$ ice around low-mass protostars. This temperature is only for surface segregation however. Therefore if more than 2~ML of pure CO$_2$ is present towards a low-mass protostar, distillation is still required. Observations show that in the sample of \citet{Pontoppidan08}, the average ice abundance of pure CO$_2$ with respect to H$_2$O ice is 2.4$\pm$2.4\%. Similar values are also found towards the low-mass and high-mass sources in the sample when calculated separately for the two types of objects. With a total ice thickness of less than 100~ML, the amount of segregated CO$_2$ towards most sources is thus consistent with surface segregation as long as the observations are dominated by ice at $>30$~K. 

To understand the origin of pure CO$_2$ ice towards individual objects requires estimates of the infall rate, the stellar luminosity and the origin of the ice absorption features  in the envelope, which is outside of the scope of this study. Such a model could however use pure CO$_2$ observations to constrain the thermal history of the protostellar envelope. While the characteristic temperatures of CO$_2$ segregation in a H$_2$O:CO$_2$ ice mixture and CO distillation from a CO$_2$:CO ice are too close to differentiate their relative contributions towards most objects, their similarity is actually an asset when using pure CO$_2$ ice as a thermal probe. Both types of ice processing are irreversible; hence once segregated or distilled, the CO$_2$ ice will remain pure.  Modeling of the amount of pure CO$_2$ ice in a protostellar envelope may then reveal the maximum size of the envelope that has been heated to 25--30~K. This inferred heating of the envelope in the past can be compared to the current luminosity of the protostar. A statistic on the discrepancy between the two would constrain the variability of protostars during an otherwise observationally inaccessible period of the stellar evolution. 

This picture may be complicated by the presence of strong UV fields around some protostars. UV radiation can produce CO$_2$ from CO and H$_2$O ices through photochemistry, photodesorb pure CO$_2$ ice, and photodissociate CO$_2$ into CO  and thus transform the pure CO$_2$ ice into a CO$_2$:CO mixture \citep{Watanabe02, Loeffler05, Oberg09b, Gerakines96}.  Of these processes, CO$_2$ formation through photochemistry should not affect the amount of pure CO$_2$ ice, since CO$_2$ then forms mixed with H$_2$O and CO ices. UV-induced destruction of some of the produced CO$_2$ ice may be important, however.  Assessing its impact for a certain object requires an individual protostellar envelope model that includes UV-induced ice processes. Without such a model, pure CO$_2$ ice observations still provides a lower limit to the amount of heat experienced by the ices in a protostellar envelope.

\section{Conclusions}

Quantifying ice dynamics is possible through a combination of a relatively large set of laboratory experiments, spanning the available parameter space, rate-equation modeling and Monte Carlo simulations. This process is more time consuming than the qualitative studies we have been building on, but has the critical advantage of providing actual rates, which can be incorporated into astrochemical models. These rates can then be used to test hypotheses about ice dynamics in space. Quantified ice processes can also be used as quantitative probes, and thus provide powerful tools to investigate processes during star formation. 

For the specific dynamical processes investigated in this study our main conclusions are:

\begin{enumerate}
\item Thin (8--37~ML) H$_2$O:CO$_2$ and H$_2$O:CO ice mixtures segregate through surface processes followed by an order of magnitude slower bulk diffusion.
\item The thin ice segregation process is qualitatively reproduced by Monte Carlo simulations where ices segregate through a combination of molecule hopping and swapping. Segregation is expected to be a general feature of ice mixtures where binding energies of all the mixture constituents are higher on average in the segregated ices compared to in the ice mixtures.
\item The surface segregation barrier is $1080\pm190$~K for the H$_2$O:CO$_2$ ice mixtures and $300\pm100$~K for H$_2$O:CO segregation.
\item H$_2$O:CO$_2$ surface segregation is ice-mixture ratio dependent, quantified as $k_{\rm seg} =2\times10^{(5\pm1)} \times e^{-1080\pm190/T}\times\left(\frac{x}{0.33}\right)^{2.5\pm1}$, where $k_{\rm seg}$ is the rate at which the top few monolayers segregate, $T$ the ice temperature and $x$ the CO$_2$ concentration in the ice with respect to the total ice amount. 
\item Segregation in thick ($>100$~ML) ices involves the bulk of the ice, which can be explained by ice cracking or H$_2$O phase transitions or both. Ice dynamics thus depends on the thickness of the investigated ice.
\item During low-mass star formation the surface segregation temperature of a typical thin H$_2$O:CO$_2$ 5:1 ice mixture is reduced to $\sim$30~K, because of the longer time scales. 
\end{enumerate}

\begin{acknowledgements}
The authors wish to thank Klaus Pontoppidan for suggesting the experiments and Ted Bergin for useful comments. Funding is provided by NOVA, the Netherlands Research School for Astronomy, a grant from the European Early Stage Training Network ('EARA' MEST-CT-2004-504604), a Netherlands Organization for Scientific Research (NWO) Veni grant and a NWO Spinoza grant.
\end{acknowledgements}

\bibliographystyle{aa}

\end{document}